\documentclass[sigconf]{acmart}
\AtBeginDocument{%
  }

\copyrightyear{2026}
\acmYear{2026}
\setcopyright{cc}
\setcctype{by}
\acmConference[MODELS 2026]{ACM/IEEE 29th International Conference on Model Driven Engineering Languages and Systems}{October 04--09, 2026}{Málaga, Spain}
\acmBooktitle{ACM/IEEE 29th International Conference on Model Driven Engineering Languages and Systems (MODELS 2026), October 04--09, 2026, Málaga, Spain}
\acmDOI{10.1145/3822455.3830313}
\acmISBN{979-8-4007-2809-9/2026/10}

\usepackage{acmart-taps}
\usepackage{multirow}
\usepackage{url}
\usepackage{algorithmic}
\usepackage{algorithm}
\usepackage{tikz}
\usepackage{ifthen}

\definecolor{DarkRed}{RGB}{137, 0, 0}
\definecolor{DarkGray}{RGB}{150, 150, 150}
\definecolor{DarkGreen}{RGB}{2, 141, 92}
\definecolor{Petunia}{RGB}{139, 0, 139}

\begin{document}

\title{ModiGen: A Large Language Model-Based Framework for Modelica Component Generation in Multi-Domain Systems}


\author{Jiahui Xiang}
\email{xiangjh@zju.edu.cn}
\affiliation{%
  \institution{Zhejiang University}
  \city{Hangzhou}
  \state{Zhejiang}
  \country{China}
}

\author{Tong Ye}
\email{tongye@zju.edu.cn}
\affiliation{%
  \institution{Zhejiang University}
  \city{Hangzhou}
  \state{Zhejiang}
  \country{China}}  
  
\author{Peiyu Liu}
\authornotemark[1]
\email{liupeiyu@zju.edu.cn}
\affiliation{%
  \institution{Zhejiang University}
  \city{Hangzhou}
  \state{Zhejiang}
  \country{China}}

\author{Yinan Zhang}
\email{zhangyinan@zju.edu.cn}
\affiliation{%
  \institution{Zhejiang University}
  \city{Hangzhou}
  \state{Zhejiang}
  \country{China}}

\author{Wenhai Wang}
\authornote{*Corresponding authors}
\email{zdzzlab@zju.edu.cn}
\affiliation{%
  \institution{Zhejiang University}
  \city{Hangzhou}
  \state{Zhejiang}
  \country{China}}
  
\renewcommand{\shortauthors}{Xiang et al.}
\begin{abstract}
Modelica is an industry-standard language for modeling and simulating complex cyber-physical systems, playing a critical role in digital twin development. Driven by the escalating demand for complex system modeling, there is a growing interest in leveraging the advanced code generation capabilities of Large Language Models (LLMs). To explore this potential, we establish a comprehensive benchmark to systematically evaluate LLMs in generating Modelica component models. Our evaluation reveals that current LLMs struggle to construct simulation-ready components due to strict acausal semantics, with failures primarily stemming from: (1) syntactic violations, (2) semantic hallucinations, and (3) physical inconsistencies.
To overcome these challenges, we propose ModiGen, a framework designed to promote syntactic correctness along with structural and physical consistency in Modelica generation. ModiGen features supervised fine-tuning to facilitate the internalization of domain knowledge. Crucially, it employs Modelica-specific structural retrieval-augmented generation to inject precise topological constraints and physical laws. Furthermore, it incorporates a fully automated, feedback-driven refinement mechanism that transforms simulation diagnostics into fine-grained instructions, guiding LLMs to resolve specific syntactic and semantic errors. Experimental results demonstrate that ModiGen achieves up to 18.65\% improvement in the functional validation pass rate over baselines, significantly enhancing the reliability of component generation.
\end{abstract}

\newcommand{\badge}[3]{
	\ifthenelse{\equal{#1}{}}{}{
		\begin{tikzpicture}[overlay, remember picture]
			\node[xshift=-5.2cm,yshift=-1.1cm] at (current page.north east) {\includegraphics[width=1.8cm]{figs/artifacts_evaluated_functional.jpg}};
		\end{tikzpicture}}
	\ifthenelse{\equal{#2}{}}{}{
		\begin{tikzpicture}[overlay, remember picture]
			\node[xshift=-3.2cm,yshift=-1.1cm] at (current page.north east) {\includegraphics[width=1.8cm]{figs/artifacts_evaluated_reusable.jpg}};
		\end{tikzpicture}}
	\ifthenelse{\equal{#3}{}}{}{
		\begin{tikzpicture}[overlay, remember picture]
			\node[xshift=-1.2cm,yshift=-1.1cm] at (current page.north east) {\includegraphics[width=1.8cm]{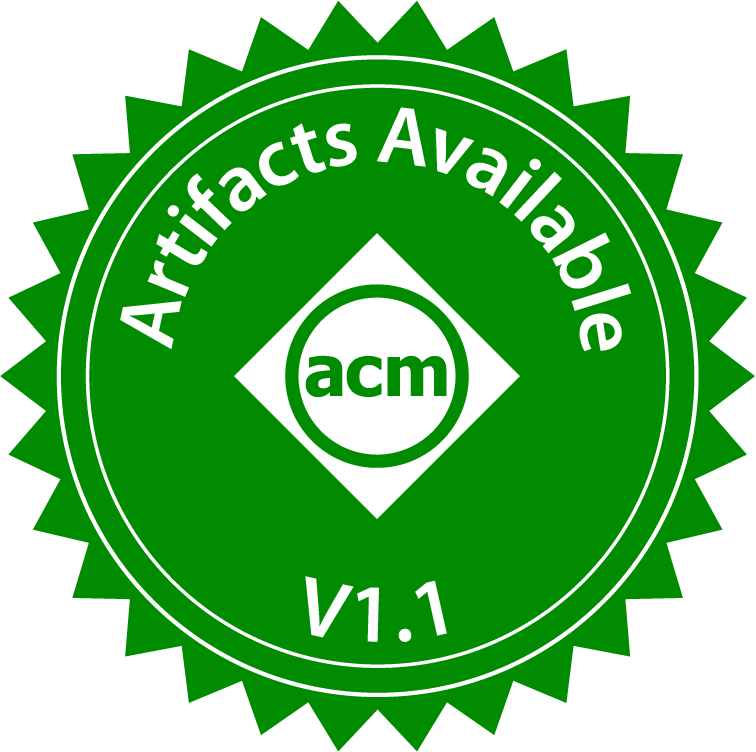}};
		\end{tikzpicture}}
}




\maketitle

\badge{}{}{available}

\section{Introduction}
With the development of Cyber-Physical Systems (CPS) and the adoption of digital twin technologies, system modeling has become a fundamental task in scientific and engineering research \cite{9715443}. Digital twins provide virtual representations of cyber-physical systems, enabling real-time simulation and monitoring. To construct and simulate such digital twins, engineers often rely on Modelica, an equation-based, non-causal, and object-oriented modeling language \cite{tiller2001introduction, fritzson2020modelica}. Modelica is widely used across domains such as mechanical \cite{zhang2024vehicle}, electrical \cite{xiao2023electrical, babaeifar2023electrical}, electronic \cite{mirz2016electronics}, hydraulic \cite{gu2018hydrostatic}, thermal \cite{sodja2008thermal, zhou2022application}, and control engineering \cite{figueiredo2014control, hirano2015vehicle}. By representing system behavior through standardized mathematical equations, Modelica provides engineers with a convenient and efficient modeling language for developing and simulating dynamic systems.

Despite its versatility, effectively using Modelica for complex system modeling remains challenging. Existing libraries often fail to cover emerging or specialized applications \cite{10.1155/2017/1578043}. Developing or adapting models demands substantial manual effort and expertise, making the process time-consuming and prone to errors.
Recent advances in large language models (LLMs) have opened new opportunities for automating code generation, including the potential to generate Modelica component models from natural language. LLMs such as OpenAI's Codex \cite{chen2021evaluating} and Meta's CodeLlama \cite{roziere2023code} demonstrate strong capabilities in code understanding and generation. However, their application to domain-specific modeling languages such as Modelica remains underexplored. This gap motivates the systematic evaluation of LLMs in generating Modelica component models (hereafter referred to as components to avoid ambiguity). Rather than relying on direct LLM-based simulation, this work focuses on Modelica component generation. Directly using LLMs to simulate physical system behaviors remains unreliable and can obscure model structures, making the results difficult to verify and maintain. Accordingly, the generated Modelica components remain explicit modeling artifacts that can be simulated, validated, reused, and maintained within established toolchains.

To this end, we propose a comprehensive benchmark designed to assess the performance of LLMs in generating Modelica components. Our evaluation reveals that current LLMs exhibit substantial limitations, with many generated components failing to simulate successfully. Specifically, these failures are primarily driven by three critical issues: (1) \textit{syntactic violations}, where generated components fail to adhere to the Modelica syntax; (2) \textit{semantic hallucinations}, where LLMs fabricate non-existent library components or parameters; and (3) \textit{physical inconsistencies}, where physical formulations violate conservation laws or interface compatibility.

To address these challenges, we propose \textbf{ModiGen}, a framework designed to enforce syntax correctness as well as structural and physical consistency in Modelica generation. The ModiGen architecture comprises three integrated stages: first, a semantic-driven fine-tuning adapts the LLM to Modelica syntax and conventions; subsequently, a Modelica-specific structural retrieval-augmented generation (RAG) module injects precise topological constraints and physical laws during generation; finally, a feedback-driven refinement module transforms raw simulation diagnostics into fine-grained instructions to rectify syntactic and semantic inconsistencies. Comparative evaluations show that ModiGen substantially outperforms baseline LLMs in both syntactic correctness and physical consistency, achieving improvements of up to 22.86\% in the simulation pass rate and 18.65\% in the functional validation pass rate.

In summary, this paper makes the following contributions:
\begin{itemize}
\item We propose a comprehensive benchmark specifically designed to evaluate the performance of LLMs in Modelica component generation.

\item We identify critical limitations in the ability of various LLMs to generate accurate and simulatable Modelica components, specifically highlighting challenges in topological and physical consistency.

\item We introduce ModiGen, a specialized framework that enhances the syntactic correctness, physical consistency, and functional reliability of Modelica component generation.

\item We empirically show that ModiGen significantly improves pass@1 over state-of-the-art LLMs, validating the necessity for domain-specific knowledge.
\end{itemize}

To facilitate future research, we make ModiGen publicly available on Zenodo at \url{https://doi.org/10.5281/zenodo.21320577}.

\section{Related Work}
This section reviews related works across three key areas: the foundational paradigms and industrial applications of Modelica, the rapid development of LLM-based code generation, and the emerging trend of automated system modeling.

\textbf{Modelica and Acausal Modeling.} Modelica serves as the industry standard for modeling complex physical systems and developing detailed digital twins. Unlike imperative programming languages such as Python or C++, which rely on causal assignment statements, Modelica employs an equation-based, acausal paradigm designed to represent physical laws directly \cite{modelica2024language} (e.g., Kirchhoff's Current Law, expressed as $i_1 + i_2 + i_3 = 0$). 
These \textit{acausal semantics} facilitate high reusability and seamless multi-domain integration, evidenced by their widespread adoption in electric vehicle systems \cite{chen2022research}, hydraulic dynamics \cite{li2016modeling}, electric power grids \cite{vanfretti2013unambiguous}, and chemical processes \cite{matejak2015free}. However, the global system of equations must be mathematically \textit{non-singular} by ensuring a precise balance between variables and equations. Furthermore, it imposes strict \textit{topological constraints}: components must be connected via compatible connector interfaces. These constraints make automated generation particularly challenging, as mere syntactic correctness does not guarantee the validity of a component.

\textbf{LLM-based Code Generation.} In recent years, AI technologies powered by LLMs have advanced rapidly, creating new opportunities for applying LLM-based code generation to a wide range of software engineering tasks \cite{ye2025problem, ye2024tram}. The Transformer architecture \cite{vaswani2017attention} laid the foundation for modern LLMs with its self-attention mechanism. While general-purpose LLMs excel at broad language understanding, code-oriented LLMs form a specialized subclass trained on extensive programming code and technical documentation. These models can perform tasks such as code generation, code comprehension, and static analysis \cite{CodeLLM2025Jiang}. Codex \cite{chen2021evaluating}, an extension of GPT-3 \cite{brown2020language}, powers GitHub Copilot, significantly enhancing code-writing efficiency. Beyond Codex, large-scale models such as AlphaCode \cite{li2022competition}, StarCoder 2 \cite{lozhkov2024starcoder}, CodeLlama \cite{roziere2023code}, Qwen2.5-Coder \cite{hui2024qwen2}, DeepSeek-Coder \cite{guo2024deepseek}, Claude 3.5 \cite{anthropic2024claude35}, and Gemini 2.5 \cite{comanici2025gemini} leverage billions of parameters and large-scale code corpora to advance automated code generation. These developments have improved the accuracy, generalization, and efficiency in this field.

\textbf{Automated Model Generation.} Beyond general code generation, LLMs have been explored for system modeling tasks. Wang et al. \cite{wang2024UML} and Cámara et al. \cite{camara2023assessment} empirically evaluated LLMs for generating Unified Modeling Language (UML) diagrams, highlighting challenges in maintaining structural integrity and consistency. Moving towards physical simulation, recent efforts have also proposed LLM-driven frameworks for digital twin modeling code generation \cite{dong2024digital}. However, the effectiveness of LLMs in generating Modelica components remains an open research question. Modelica components represent dynamic behaviors governed by differential-algebraic equations. Consequently, the generated component must satisfy both syntax and domain-specific physical laws \cite{joel2024survey}. Despite grasping basic syntax, LLMs frequently exhibit semantic hallucinations—fabricating non-existent components and breaking the topological consistency required for large-scale systems. The challenge aligns with broader difficulties in code generation for Low-Resource Programming Languages (LRPLs) and Domain-Specific Languages (DSLs) \cite{GeoFub2025Hou}.

\section{Benchmark and Evaluation}
As system modeling expands into emerging domains, existing standard libraries often lack specialized components \cite{10.1155/2017/1578043}. Consequently, our research task investigates how LLMs can automate the generation of custom Modelica components to satisfy specific modeling requirements, thereby enhancing multi-domain adaptability and design efficiency. To assess this capability, we establish a comprehensive evaluation framework. This section details the benchmark dataset design, the formulation of the evaluation framework, and the baseline results obtained from representative LLMs.

\subsection{Benchmark Datasets}
Currently, no universally accepted benchmark exists for evaluating LLM-based generation of Modelica components. To bridge this evaluation gap and ensure a focused assessment across diverse physical domains, we constructed a specialized benchmark leveraging the rich ecosystem maintained by the Modelica Association and the open-source community \cite{modelica_libraries}. The resulting dataset is a hybrid corpus consisting of 126 high-quality components: 105 components were meticulously selected from industrial-grade ecosystems, while 21 components were specifically synthesized by the authors.

\begin{figure}
\centering
  \setlength{\abovecaptionskip}{1mm}
  \setlength{\belowcaptionskip}{-4mm}
\includegraphics[width=\linewidth]{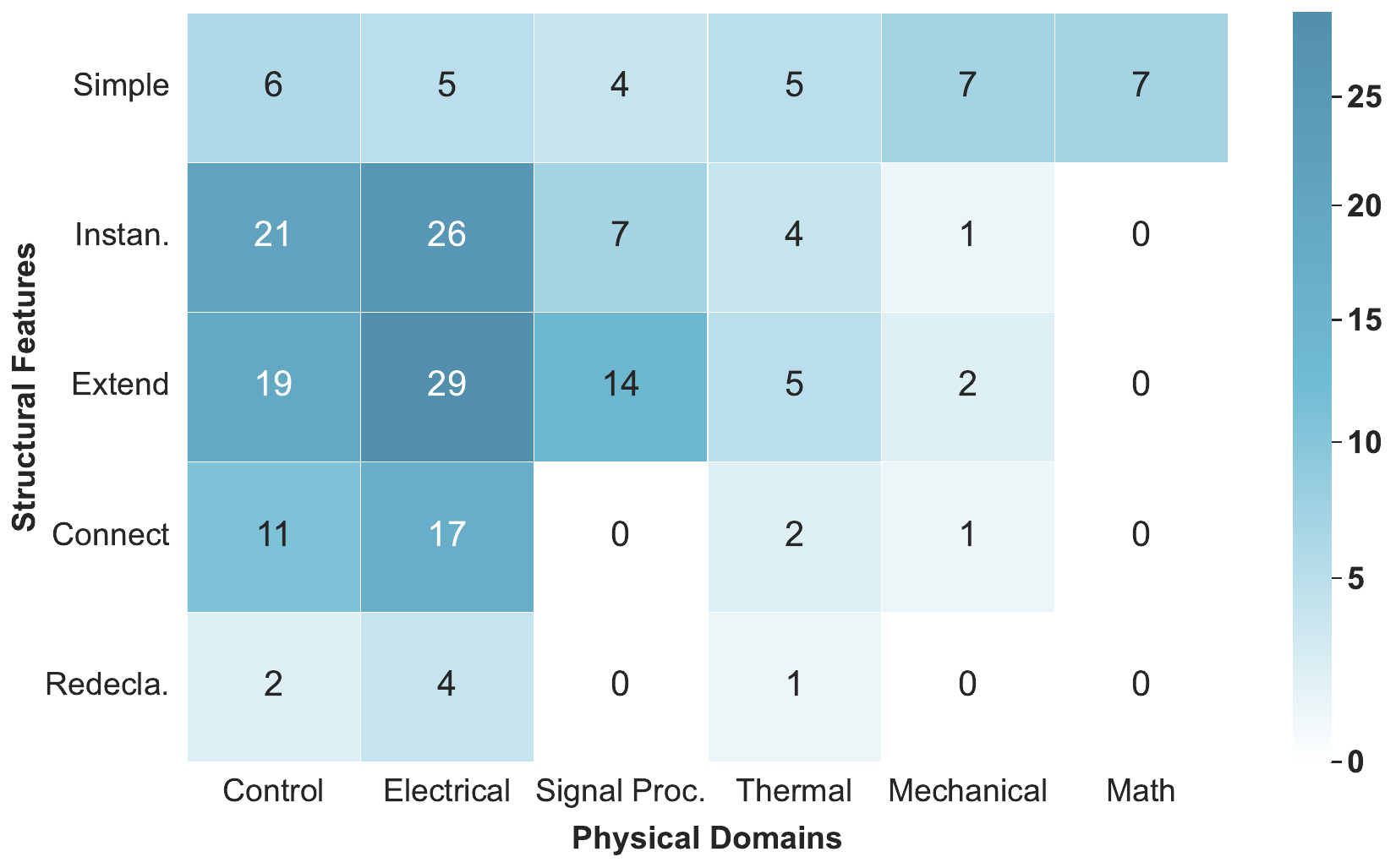}
\caption{Distribution of the Benchmark Dataset Across Primary Physical Domains and Structural Complexities.}
\Description{}
\label{fig:1}
\end{figure}

To ensure a comprehensive evaluation reflecting real-world industrial applications, our benchmark dataset is primarily tailored toward CPS. Consequently, the physical domain of the dataset predominantly features components from control systems, electrical engineering, and signal processing. In addition, the dataset encompasses fundamental components from thermodynamics, mechanics, mathematics, biochemistry, and fluid mechanics to evaluate broader multi-domain capabilities. Regarding structural complexity, the benchmark incorporates varying difficulty tiers related to Modelica syntax—ranging from basic algebraic equations (\textit{Simple}), to component instantiation (\textit{Instan.}), topological interface connections (\textit{Connect}), object-oriented inheritance (\textit{Extend}), and advanced class redeclarations (\textit{Redecla.}). The detailed distribution of these components across the physical domains and structural complexities is illustrated in Figure \ref{fig:1}. Following the specialized component classes defined in the Modelica \cite{modelica2024language}, our dataset comprises 91 \texttt{model}s, 21 \texttt{block}s, 13 \texttt{function}s, and 1 \texttt{class}, facilitating a comprehensive assessment of acausal and structured modeling capabilities.

To transform these raw components into a standardized benchmark dataset for automated evaluation, we implemented a systematic data processing pipeline. First, we traversed the \texttt{.mo} files and extracted component-level content together with relevant metadata, including library source, Modelica version, documentation, and hierarchical source information. This metadata helps preserve the organizational context of each component and supports subsequent prompt construction and evaluation. Second, we performed source code cleaning by removing \texttt{annotation} blocks---elements primarily used for GUI rendering---as well as non-modeling content, redundant whitespace, and unnecessary line breaks. This step reduces syntactic noise while retaining the functional logic and structural information required for component generation. We further applied path-based deduplication and similarity filtering to remove overlapping or highly similar components. Finally, to simulate realistic modeling scenarios, we constructed natural-language prompts based on the cleaned code and documentation for each component. To ensure the benchmark's empirical rigor, all test prompts underwent manual verification to confirm that they accurately reflect the physical intent and structural requirements of the target components.

\subsection{Evaluation Framework}\label{sec:Metrics}
To comprehensively evaluate LLMs' performance in Modelica component generation, we design an execution-based evaluation framework that assesses both syntactic correctness and functional reliability. We use OpenModelica \cite{fritzson2020openmodelica}, a widely adopted open-source simulation environment for Modelica, to simulate the generated components. 
The evaluation procedure consists of four sequential steps: 

\textbf{(a) Loading verification} verifies that the generated component conforms to the basic syntax and lexical structure of Modelica.

\textbf{(b) Consistency verification} inspects the component for errors such as incorrect parameter types, missing component invocations, or violations of structural constraints (e.g., unbalanced equations and variables).

\textbf{(c) Simulation verification} ensures that the component can simulate successfully, without issues such as uninitialized variables or algebraic loops. For \texttt{function} components, we construct appropriate use cases for verification.

\textbf{(d) Function validation} compares the aligned simulation output signal $y^V$ with an expected reference trajectory $y^R$. Taking transient behavior and physical consistency into account, we consider the generation valid if the root mean square error (RMSE) falls below a threshold $\varepsilon$. For a reference sequence of length $T_r$, the validation condition is defined as:
$\mathrm{RMSE} = \sqrt{\frac{1}{T_r}\sum_{i=1}^{T_r} \bigl(y^V_i- y^R_i\bigr)^2} < \varepsilon.$

To avoid false positives, outputs passing RMSE-based functional validation are further examined by two independent blind annotators, who reject components with physical or logical inconsistency regardless of RMSE.

Correctly implemented components typically yield consistent simulation behavior, though minor deviations may arise from differences in solvers or simulation step sizes. To account for such computational noise, we adopt the empirically determined tolerance of $\varepsilon = 3.16\times10^{-2}$. The threshold is sufficiently larger than solver-induced numerical errors while remaining strict enough to flag functionally incorrect behaviors. When simulation steps differ from the reference component, time alignment is performed before validation. After alignment, the validation process identifies the output channel that best corresponds to the reference output. The pointwise deviation between the two channels is then computed over the full simulation horizon using the RMSE metric. The overall functional validation workflow is summarized in Algorithm~\ref{alg:alg1}.

\begin{algorithm}\small
\caption{Function Validation via Time-Aligned RMSE}\label{alg:alg1}
\begin{algorithmic}
\STATE\textbf{Input:} Reference trajectory $R\in\mathbb{R}^{T_r\times (1+m)}$, validated trajectory $V\in\mathbb{R}^{T_v\times (1+n)}$. (First column is time, remaining columns are signals) \\
\STATE\textbf{Output:} PASS/FAIL validation decision. \\[2pt]
\STATE 1: Extract time vectors $t^R$, $t^V$ and corresponding signal vectors.
\STATE 2: \textbf{Align time grids:} For each $t^R_i$, select the closest timestamp in $t^V$ (interpolation optional if $T_r > T_v$) to construct an aligned trajectory $\tilde V$.
\STATE 3: \textbf{Select best-matching signal:} Among all candidate output channels in $\tilde V$ (columns $\ell$), choose the one that best matches the reference output signal $y^R$:
\[
\ell^* = \arg\min_\ell \sqrt{\frac{1}{T_r} \sum_{i=1}^{T_r} (y^R_i - \tilde V_{i,\ell})^2}
\]
\STATE 4: \textbf{Compute RMSE:} 
\STATE \quad Let the selected output signal be $y^V_i = \tilde V_{i,\ell^*}$.
\[
\mathrm{RMSE}=\sqrt{\frac{1}{T_r}\sum_{i=1}^{T_r} (y^R_i - y^V_i)^2}
\]
\STATE 5: \textbf{Decision:}
\STATE \quad \textbf{if} $\mathrm{RMSE}<\varepsilon$ \textbf{then} PASS; \textbf{else} FAIL.
\end{algorithmic}
\end{algorithm}

This validation evaluation is highly stringent: if a generated component fails at any stage, it is classified as non-compliant. To ensure scalability, the process is \textbf{fully automated} (with the exception of manual inspection), with the first three structural steps seamlessly executed via the OMPython interface in OpenModelica \cite{OMPython2025}.

\textbf{Evaluation Metrics.} For quantitative evaluation, we adopt the widely used $pass@k$ as the metric \cite{chen2021evaluating}. To precisely diagnose the specific failure modes of LLMs, we decompose the overall pass rate into four stage-specific variations corresponding to our evaluation pipeline: $pass_{load}@k$ (pass loading verification), $pass_{cons}@k$ (pass consistency verification), $pass_{sim}@k$ (pass simulation verification), and ultimately $pass_{func}@k$ (pass function validation). This granular design provides a multidimensional evaluation, effectively pinpointing LLM capabilities across distinct types of modeling challenges. Given that Modelica component generation is an exceptionally complex task requiring the synthesis of syntax, logic, and multi-domain physical laws, we evaluate $pass@1$ and $pass@10$. For the multiple generation attempts, we set $n=10$ samples per task to strike a balance between statistical reliability and computational overhead.

\subsection{Evaluation of Baselines}
To establish a comprehensive baseline for the generation task, we evaluate six representative LLMs. For open-weights models, we select a leading general-purpose model, LLaMA-3.1-8B-Instruct (abbreviated as LM3.1-8B) \cite{grattafiori2024llama}, alongside three state-of-the-art code-specialized models: DeepSeek-Coder-7B-Instruct-V1.5 (DSC1.5-7B), Qwen2.5-Coder-7B-Instruct (QWC2.5-7B), and Qwen2.5-Coder-32B-Instruct (QWC2.5-32B) \cite{hui2024qwen2}. Furthermore, to establish an upper bound on current generation capabilities, we incorporate leading proprietary frontier models: GPT-4o \cite{openai2024gpt4o} and Claude-3.5 Sonnet (Claude-3.5) \cite{anthropic2024claude35}. Regarding hyperparameter configurations, we employ a sampling temperature of 0.3 to strike an optimal balance between maintaining syntactical stability and allowing diversity for valid solutions. Additionally, the top-$k$ sampling parameter is fixed at 10, effectively preserving semantic expressiveness while ensuring generation reliability.

\begin{table}
\centering
  \setlength{\abovecaptionskip}{1mm}
  \setlength{\belowcaptionskip}{0mm}
\caption{\label{tab:1}Performance of LLMs on Modelica Generation.}
\scalebox{0.85}{
\begin{tabular}{l|cccc}
\toprule
LLM (\%) & $pass_{load}@1$ & $pass_{cons}@1$ & $pass_{sim}@1$ & $pass_{func}@1$\\
\midrule
DSC1.5-7B & 74.21 & 29.84 & 14.21 & 9.44 \\
LM3.1-8B & 61.11 & 20.63 & 10.08 & 6.11\\
QWC2.5-7B & 54.37 & 26.75 & 13.65 & 9.05 \\
QWC2.5-32B & 69.29 & 31.59 & 19.29 & 18.15 \\
GPT-4o & \textbf{95.56} & \textbf{50.16} & \textbf{27.38} & \textbf{24.44} \\
Claude-3.5& 88.41 & 44.44 & 26.27 & 23.57 \\ 
\midrule
LLM (\%) & $pass_{load}@10$ & $pass_{cons}@10$ & $pass_{sim}@10$ & $pass_{func}@10$\\
\midrule
DSC1.5-7B & 93.65 & 46.83 & 28.57 & 19.84 \\
LM3.1-8B & 88.89 & 42.06 & 22.22 & 14.29 \\
QWC2.5-7B & 77.78 & 44.44 & 26.98 & 16.67 \\
QWC2.5-32B & 96.83 & 54.76 & 35.71 & 29.37 \\
GPT-4o & 99.21 & 64.29 & 38.10 & 34.13 \\
Claude-3.5& \textbf{100.0} & \textbf{67.46} & \textbf{43.65} & \textbf{40.48} \\
\bottomrule
\end{tabular}}
\end{table}

Table~\ref{tab:1} summarizes LLM performance on the generation task. Proprietary LLMs significantly outperform smaller open-weight models across all metrics. Across each evaluation stage, GPT-4o achieves the highest performance at $pass@1$, while Claude-3.5 exhibits the highest overall performance at $pass@10$. Among the open-weight models, DSC1.5-7B and QWC2.5-7B slightly outperform LM3.1-8B. This suggests that code-centric pretraining still provides a slight advantage in comprehending structured modeling languages. QWC2.5-32B achieves the superior performance among all open-weight candidates. This result aligns with the broader scaling trend in LLMs, where an increased parameter count typically correlates with enhanced generation capabilities. Most importantly, a consistent performance reduction is observed across the four evaluation stages: the pass rates shrink progressively from $pass_{load}@k$ to $pass_{cons}@k$, to $pass_{sim}@k$, and finally to $pass_{func}@k$. This reflects the multifaceted challenges LLMs encounter during generating Modelica components, such as lexical and syntactic violations (\textit{load} failure), invalid inheritance or connections (\textit{cons} failure), singularities or numerically unsolvable equations (\textit{sim} failure), and deviations from expected dynamic trajectories (\textit{func} failure).

Consequently, directly prompting LLMs to generate valid Modelica components is often insufficient. This underscores the necessity for further enhancement strategies to improve their adaptability and reliability in complex system modeling tasks.

\section{ModiGen}
To address the limitations identified in the baseline generation process, we propose \textbf{ModiGen}, an innovative generation framework strictly tailored for the Modelica language. The core architecture and the specific methods of this framework are detailed below.

\subsection{Framework Architecture}
\begin{figure}
\centering
  \setlength{\abovecaptionskip}{1mm}
  \setlength{\belowcaptionskip}{-4mm}
\includegraphics[width=\linewidth]{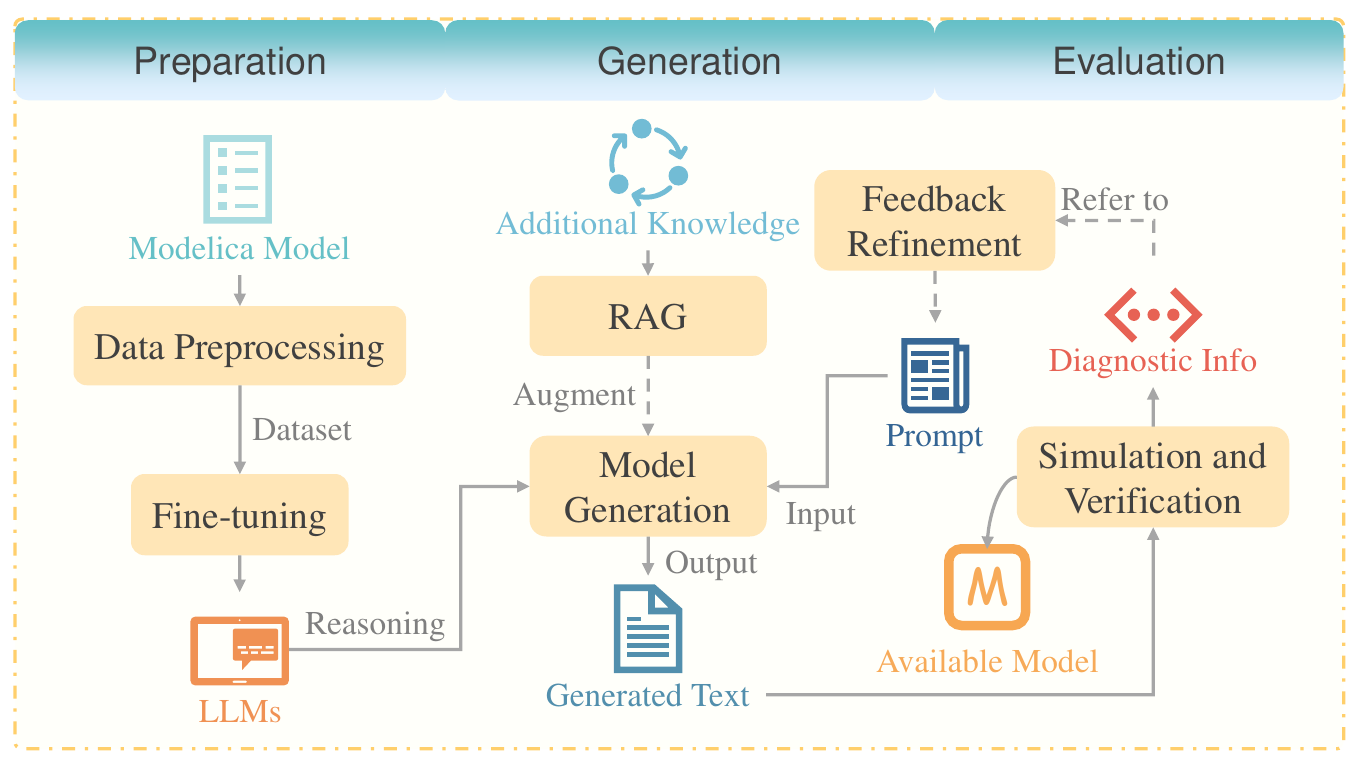}
\caption{Overview of the ModiGen Architecture.}
\Description{}
\label{fig:2}
\end{figure}

The framework architecture of ModiGen, illustrated in Figure \ref{fig:2}, is designed as a generation pipeline that integrates domain knowledge with an automated simulation module. Initially, the framework constructs a foundational knowledge base by preprocessing industrial Modelica components. Leveraging the component code and associated prompts from this knowledge base, the LLM undergoes domain-specific fine-tuning to explicitly internalize the specialized syntax and physical semantics. During the generation phase, this fine-tuned LLM operates in collaboration with an optional Modelica-specific structural RAG module. The RAG module dynamically retrieves semantic and relational context, enabling the LLM to more effectively capture domain-specific expertise and associated topological structures. Finally, to ensure adherence to Modelica's rigorous standards, each component undergoes automated compilation and function validation. If any violation is detected, precise error feedback is routed back to the LLM for further refinement.

\subsection{Knowledge Internalization}\label{sec:template}
To effectively internalize unique syntactic structures and acausal physical semantics of Modelica, ModiGen relies on a highly curated, domain-specific training pool. While the Modelica community possesses extensive libraries, these repositories are decentralized and structurally noisy. To establish a standardized foundational knowledge base, we initially survey a broad pool of 44 open-source libraries. However, to ensure strict alignment with our benchmark, we avoid indiscriminately aggregating all available files. Instead, we implement a two-step curation strategy: purification and selective extraction. We first apply a rigorous data purification pipeline to these files, filtering out samples that lack descriptive documentation, exceed optimal context thresholds, or contain non-modeling graphical metadata (e.g., \texttt{UserGuide}, \texttt{Icon}). We then selectively extract the specific modeling components that align with the multi-domain evaluation tasks established in our benchmark. 

To seamlessly align the LLMs with human generation intent, we directly construct a specialized instruction fine-tuning dataset from these purified components. Specifically, we leverage the GPT-4 API to reverse-generate a natural language prompt for each filtered Modelica component. To adhere to a standardized prompt format consistent with the benchmark evaluation, we develop a standardized template that ensures each instance encapsulates the following critical dimensions: (1) \textbf{Component Functionality}: A detailed specification of the component's intended purpose and operational logic. (2) \textbf{Parameters}: Explicit definitions of data types, default values, and physical units for all inputs, outputs, and constants. (3) \textbf{Laws and Principles}: Identification of core governing laws or high-level physical principles that guide the LLM in deriving valid mathematical formulations. (4) \textbf{Inheritance and Interfaces}: A specification of the model's architectural hierarchy, detailing the inheritance from base components and the parameterization of standardized interfaces for modular connectivity. This multi-dimensional structure provides a rigorous framework for capturing task requirements and physical constraints, ensuring that the fine-tuned LLMs internalize the deep semantics of Modelica modeling. 

To illustrate the prompt template, Table \ref{tab:sft_sample} presents a representative example. After quality assurance and de-duplication, this synthesis yields an instruction-supervised dataset, $Dataset_{sft}$. It comprises 3,196 cases, mainly consisting of 2,268 \texttt{model}s, 251 \texttt{block}s, 260 \texttt{partial model}s, and 261 \texttt{function}s, along with 156 instances of other types such as \texttt{record} and \texttt{connector}. This dataset ensures a balanced representation of Modelica's structural complexities without introducing statistical bias. Crucially, to prevent data contamination, we enforce a strict exclusivity: no sample from the benchmark dataset is included in this training pool.

\begin{table}
\centering
\small
  \setlength{\abovecaptionskip}{1mm}
  \setlength{\belowcaptionskip}{0mm}
\caption{An Instruction-Tuning Sample Utilizing the Standardized Prompt Format.}
\label{tab:sft_sample}
\begin{tabular}{p{0.95\linewidth}}
\toprule
\textbf{Field - Content}\\
\midrule
\textbf{Instruction} - You are an expert in modeling with Modelica. Please generate the corresponding model based on the following description. \\
\midrule
\textbf{Input} - Develop a Modelica model for an ideal linear resistor with a Real parameter $R$ (default 1 Ohm). The model must utilize standard Pin connectors $p$ and $n$ from `CustomLib.Electrical.Analog.Interfaces'. Ensure the physical behavior is governed by Kirchhoff’s Current Law and Ohm's Law. \\
\midrule
\textbf{Output} -\newline
\begin{imageonly} 
\begin{verbatim}
model Resistor "Ideal linear electrical resistor"
  parameter Real R = 1 "Resistance in Ohms";
  CustomLib.Electrical.Analog.Interfaces.Pin p;
  CustomLib.Electrical.Analog.Interfaces.Pin n;
equation
  p.i + n.i = 0; p.v - n.v = R * p.i;
end Resistor;
\end{verbatim}\end{imageonly}
\\
\bottomrule
\end{tabular}
\end{table}

For the knowledge internalization, we employ Low-Rank Adaptation (LoRA) \cite{hu2022lora} on the leading open-weights LLMs identified in our baseline evaluation. This Parameter-Efficient Fine-Tuning (PEFT) approach \cite{lialin2023scaling} injects domain-specific physical semantics into the LLMs while mitigating catastrophic forgetting of their foundational reasoning capabilities. During the instruction fine-tuning phase, the optimization is driven by a cosine learning rate scheduler with an initial rate of 1e-4 and a batch size of 1 per device. Training is strictly capped at 10 epochs to ensure stable convergence and robust generalization across diverse modeling tasks. The adapter parameters are selected based on the checkpoint that minimizes evaluation loss, thereby ensuring robust generalization and mitigating the risk of overfitting to the specific fine-tuning distribution.

\subsection{Knowledge Augmentation}
Generating a valid Modelica component requires more than semantic understanding; it demands strict adherence to both intricate topological structures and rigorous physical principles. While fine-tuning effectively equips the LLM with basic programming patterns, it often struggles to reliably preserve global structural consistencies and memorize complex mathematical formulations \cite{gekhman2024does, ghosh2024closer}. Consequently, augmenting the generation process with external domain knowledge becomes imperative. However, traditional Retrieval-Augmented Generation (RAG) \cite{10.5555/3495724.3496517} relies on linear textual chunking, which inevitably fragments these vital structural dependencies \cite{mavromatis2024gnn}. To address this limitation, we introduce a \textbf{Modelica-specific structural RAG} mechanism. By parsing and organizing domain knowledge according to the language's inherent syntactic hierarchy and object-oriented boundaries, this approach enables the generative LLM to retrieve intact topological structures and cohesive physical principles, rather than isolated textual snippets.

We construct a domain-specific structural knowledge base to encode the multi-domain knowledge and structural dependencies inherent in Modelica. Instead of flattening the source code into continuous text, we organize the reference corpus ($Corpus_{rag}$) into discrete structural entities (e.g., \texttt{model}, \texttt{function}, \texttt{connector}). As defined by object-oriented characteristics of Modelica, each entity is encapsulated alongside its fully qualified hierarchical path, a concise functional description, and detailed technical documentation. The namespace hierarchy enhances the accuracy of path dependencies required for cross-component inheritance, instantiation, and invocation during component generation, whereas the descriptions and technical documentation establish a robust foundation for the requisite domain knowledge and governing physical laws.

\textbf{Structural Index Construction.} To construct the structural knowledge index, we process this pre-structured corpus using a retrieval framework equipped with a multilingual embedding model, such as BGE-M3 \cite{chen2024bge}. To prevent the fragmentation of vital structural dependencies, we implement an extended chunking strategy with a 2,048-token window. This approach preserves the structural integrity of complex Modelica components across both their physical equations and topological interfaces. Crucially, we introduce a \textit{retrieval-generation context decoupling} strategy to maximize the signal-to-noise ratio during semantic search. Under this strategy, the embedding process is restricted to raw source code, hierarchical paths, and concise descriptions, mapping core functional features into the vector space. Verbose natural language documentation is explicitly excluded to prevent semantic dilution. Upon successful retrieval, this excluded documentation is re-integrated into the LLM's prompt, providing specialized context for generation.

\textbf{Retrieval-Augmented Generation.} 
During the generation phase, the vector store acts as a dynamically searchable knowledge base. When the LLM requires external context, it formulates a query to traverse the structural index. Our mechanism extracts metadata-enriched text chunks, which encapsulate the target component along with its extended structural dependencies, such as required interfaces and physical constraints. The retrieved content is then serialized into the prompt, guiding the LLM to generate valid equations and connections. We configure our retrieval framework to dynamically fall back to standard non-augmented generation if no highly relevant context is identified, thereby mitigating performance regression induced by retrieval noise.
Furthermore, to avert \textit{knowledge conflict} between the SFT memory and RAG guidance for open-source LLMs and achieve optimal performance, we introduce a \textit{conditional routing mechanism}: based on the dynamic generation status—specifically, the relevance of the retrieved context meeting a predefined threshold—the framework invokes the RAG module on demand to overcome domain-knowledge deficits.

\begin{figure}[h]
\centering
  \setlength{\abovecaptionskip}{0mm}
  \setlength{\belowcaptionskip}{-4mm}
\includegraphics[width=\linewidth]{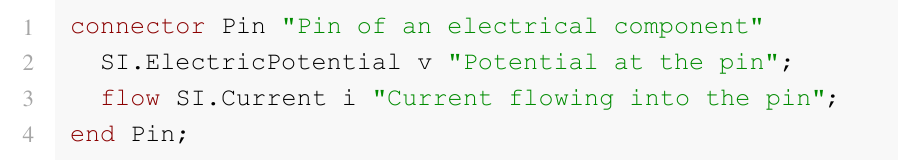}
\caption{Definition of the \texttt{Pin} connector.}
\Description{}
\label{lst:2}
\end{figure}

\begin{figure}[h]
\centering
  \setlength{\abovecaptionskip}{1mm}
  \setlength{\belowcaptionskip}{-4mm}
\includegraphics[width=\linewidth]{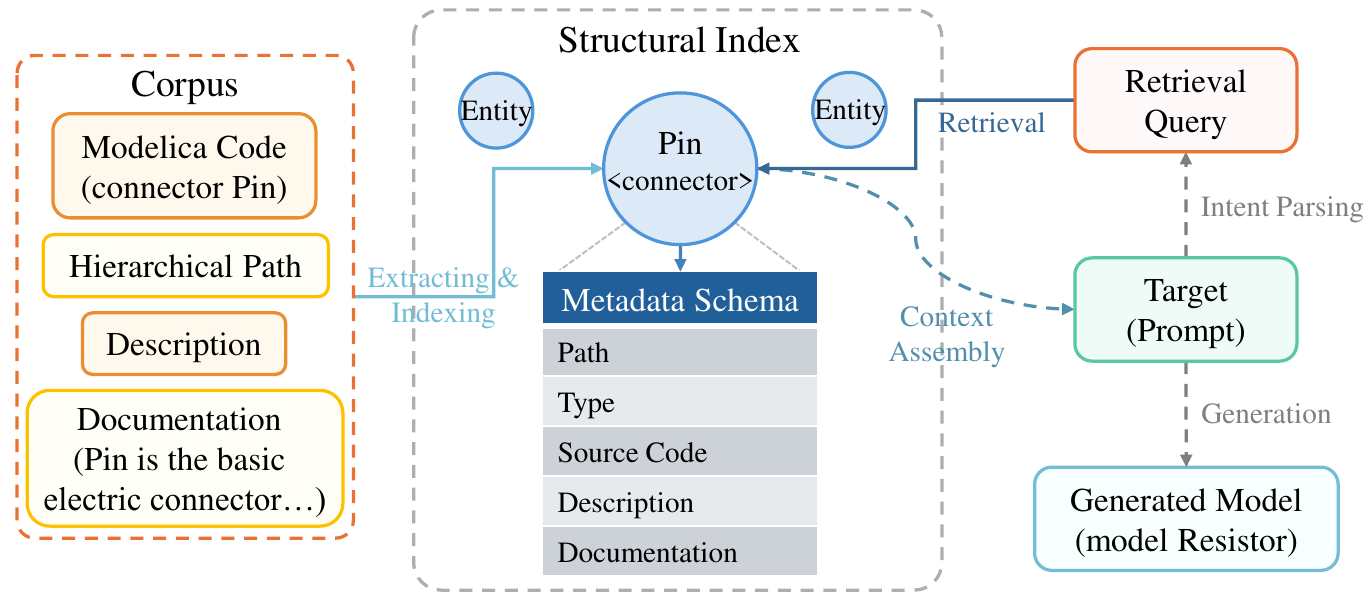}
\caption{An Illustrative Example of Knowledge Augmentation for Interdependent Component Generation.}
\Description{}
\label{fig:3}
\end{figure}

\textbf{Illustration Example.} To illustrate how knowledge augmentation provides crucial support during generation, we present a foundational dependency case: consider the scenario where the LLM needs to generate an ideal electrical resistor shown in Table \ref{tab:sft_sample}. To formulate equations for the \texttt{Resistor}, the LLM must understand the specific physical variables associated with its required \texttt{Pin} connector, as shown in Figure~\ref{lst:2}. During the index construction phase, the extraction engine parses the reference corpus to identify functional components. For the \texttt{Pin} connector, the engine encodes its source code and the related information into a unified vector representation within the structural index. Simultaneously, its internal variables, \texttt{v} and \texttt{i}, along with their detailed technical documentation, are preserved as associated metadata. Consequently, when generating the \texttt{Resistor}, the LLM retrieves this metadata-enriched entity, enabling it to formulate equations based on the precise context of the \texttt{Pin}. The overall flow of this process is illustrated in Figure~\ref{fig:3}. This structural and semantic fusion directly supports the accurate generation of these interdependent components.


\subsection{Feedback-Driven Refinement}
Ensuring correctness requires not only high-quality initial generation but also an effective mechanism for refining the component based on validation results. Existing research, such as OpenCodeInterpreter \cite{zheng-etal-2024-opencodeinterpreter} and RLTF \cite{liu2023rltf}, has demonstrated that feedback-driven reinforcement through dynamic testing can enhance code quality. These approaches rely on verifying the correspondence between the generated code and its execution results, using deterministic feedback from compilation or simulation environments to guide iterative refinement of the generated code \cite{shinn2023reflexion, chen2024teaching}. In this study, we introduce an automated, feedback-driven refinement mechanism tailored for Modelica's rigorous compilation pipeline.

We begin by applying an automated cleaning procedure to post-process the generated text, ensuring basic syntactic compliance. Crucially, before importing the generated component into the simulation environment, we pre-load the requisite libraries that contain its structural dependencies. This step ensures environment consistency and isolates genuine modeling errors from mere missing dependencies. Once the library dependencies are prepared, the cleaned Modelica text is imported into the environment for detailed evaluation, as described in Section \ref{sec:Metrics}.

During the sequential phases of loading, consistency checking, and dynamic simulation, any verification failures trigger specific compiler-generated diagnostics. To ensure these diagnostics are actionable for the LLM, our framework processes them into highly granular feedback. This information captures errors at the \textbf{token level}, ranging from superficial lexical omissions to structural modeling violations, and ultimately to runtime numerical exceptions. Furthermore, it pinpoints these errors with precise line and column positioning, thereby allowing the LLM to locate and rectify specific syntactic violations efficiently. These exact diagnostics, coupled with tailored refinement instructions, are then fed back into the LLM along with the original prompt and the current component iteration to guide targeted corrective revisions. If the component passes compilation but fails functional validation, feedback is provided to highlight specific discrepancies between the simulation results and the expected physical behaviors.

\begin{figure}
\centering
  \setlength{\abovecaptionskip}{1mm}
  \setlength{\belowcaptionskip}{-4mm}
\includegraphics[width=\linewidth]{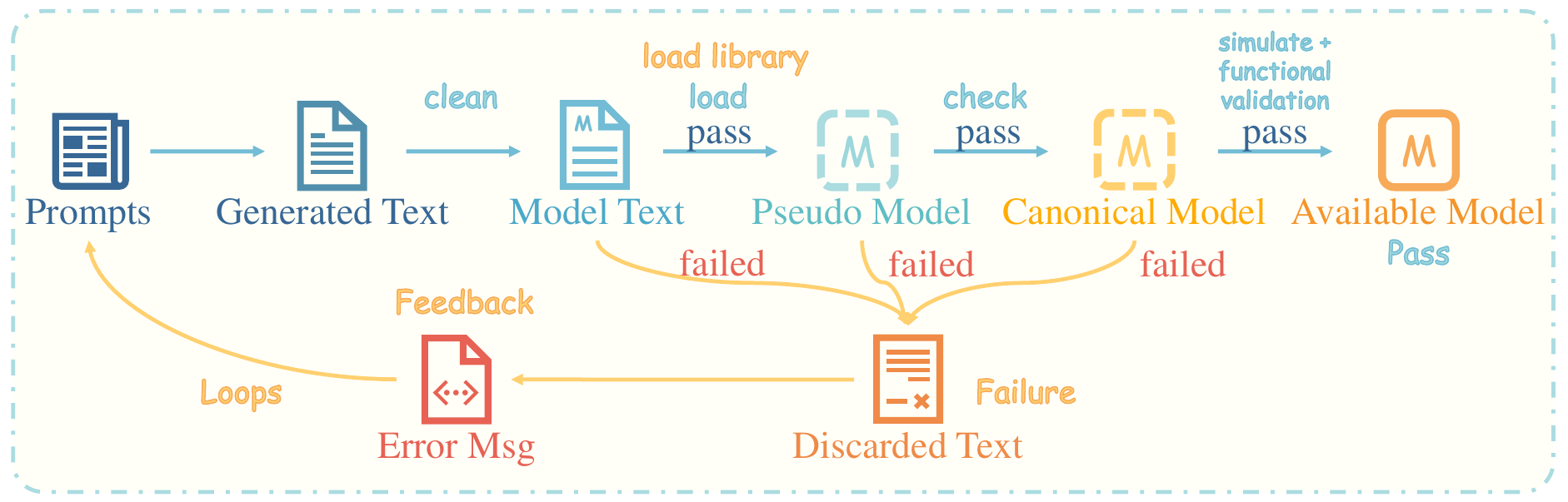}
\caption{The Flowchart of Modelica Generation and Feed-back-driven Refinement Mechanism.}
\Description{}
\label{fig:4}
\end{figure}

By applying this feedback-driven refinement, the LLM is guided toward generating components that satisfy both valid syntax and functional specifications. The complete flowchart of Modelica generation and feedback-driven refinement is shown in Figure \ref{fig:4}.

\section{Comprehensive Evaluation}
We formulated three research questions (\textbf{RQs}) to systematically evaluate the performance of ModiGen:

\textbf{RQ1:} To what extent does ModiGen improve the generation performance of LLM compared to the baseline?

\textbf{RQ2:} How do different enhancement strategies individually impact the quality of generated Modelica components?

\textbf{RQ3:} What is the computational overhead introduced by our framework when generating Modelica components?

This section addresses these RQs through empirical evaluation, offering detailed insights into the effectiveness and efficiency of ModiGen. All experiments involving open-source LLMs were conducted on a server equipped with four NVIDIA A100 (80GB) GPUs.

\subsection{Evaluation of ModiGen}
In response to \textbf{RQ1}, we conducted a series of experiments to evaluate the effectiveness of ModiGen systematically. We quantified the overall performance improvements by comparing the application of the framework to the baseline, specifically measuring $pass_{sim}@k$ and $pass_{func}@k$ ($k$ = 1, 10) to demonstrate how ModiGen enhances Modelica component generation. While $pass_{sim}@k$ verifies the simulability, $pass_{func}@k$ serves as the ultimate criterion for success. Notably, our framework is applicable to both open-source and proprietary LLMs. For closed-source models, such as GPT-4o and Claude-3.5, the fine-tuning step is omitted due to the unavailability of model weights; whereas open-source LLMs can fully leverage the framework, including fine-tuning, to achieve maximum performance gains. Table \ref{tab:2} presents the experimental results using our framework ModiGen (denoted as \textbf{MG}). The detailed analysis reveals three core insights:

\begin{table}
\centering
\setlength{\abovecaptionskip}{1mm} 
\setlength{\belowcaptionskip}{1mm}
\caption{\label{tab:2}Performance Enhancement of LLMs for Modelica Generation with ModiGen}
\scalebox{0.82}{
\begin{tabular}{l|cccc}
\toprule
LLM (\%) & $pass_{sim}@1$ & $\Delta pass_{sim}@1$ & $pass_{func}@1$ & $\Delta pass_{func}@1$\\
\midrule
DSC1.5-7B & 27.86 & +13.65 & 17.94 & +8.50 \\
LM3.1-8B & 32.94 & +22.86 & 24.76 & \underline{+18.65} \\
QWC2.5-7B & 26.98 & +13.33 & 21.98 & +12.93 \\
QWC2.5-32B	& 42.62 & \underline{+23.33} & 33.17 & +15.02 \\
GPT-4o & \textbf{44.60} & +17.22 & \textbf{36.90} & +12.46 \\
Claude-3.5 & 42.54 & +16.27 & 36.51 & +12.94 \\
\midrule
LLM (\%) & $pass_{sim}@10$ & $\Delta pass_{sim}@10$ & $pass_{func}@10$ & $\Delta pass_{func}@10$\\
\midrule
DSC1.5-7B & 44.44 & +15.87 & 28.57 & +8.73 \\
LM3.1-8B & 57.14 & \underline{+34.92} & 38.89 & \underline{+24.60} \\
QWC2.5-7B & 42.86 & +15.87 & 31.75 & +15.08 \\
QWC2.5-32B & 65.87 & +30.16 & 42.06 & +12.70 \\
GPT-4o & \textbf{58.73} & +20.63 & \textbf{50.79} & +16.67 \\
Claude-3.5 & 57.94 & +14.29 & 48.41 & +7.94 \\
\bottomrule
\end{tabular}}
\end{table}

\textbf{General Applicability.} 
The integrated framework delivers performance gains across various model architectures and parameter scales, consistently demonstrating its effectiveness. Notably, proprietary LLMs, despite omitting fine-tuning, still achieve improvements through the RAG and feedback-driven refinement modules, highlighting the contributions of these modules to structured dependency retrieval and refinement. For instance, GPT-4o achieves the highest overall performance, with $pass_{sim}@1$ reaching 44.60\% (+17.22\%) and $pass_{func}@1$ reaching 36.90\% (+12.46\%). Furthermore, the framework shows scalability with respect to model capacity. As the base reasoning capabilities of the LLMs increase (e.g., transitioning from the 7B to the 32B scale of the Qwen2.5-Coder family), they demonstrate an ability to better utilize the end-to-end generative pipeline. Overall, the current results indicate that all LLMs that utilize ModiGen have achieved effective performance improvements in addressing Modelica component generation.

\textbf{Synergistic Enhancement.} 
The generating Modelica components should be simulatable and functionally available. The results indicate that our method successfully addresses both dimensions concurrently. Rather than merely acting as a syntax corrector, the framework incorporates comprehensive domain knowledge and physical constraints to drive logical accuracy. Notably, LM3.1-8B achieves a 18.65\% increase in $pass_{func}@1$, mirroring its 22.86\% growth in $pass_{sim}@1$. This synergistic enhancement confirms that the comprehensive framework enables LLMs to effectively capture both the structural rules and the physical logic of the target domain.

\textbf{Benefits of Multiple Candidates.} 
As the generation attempts increase from 1 to 10, the performance improvements become more substantial. For example, in terms of QWC2.5-32B, the $pass_{sim}@10$ increases by 30.16\% and the $pass_{func}@10$ also rises by 12.70\%. This collective improvement allows lighter open-source LLMs to perform competitively with proprietary LLMs. Fundamentally, this behavior indicates that ModiGen successfully mitigates the inherent randomness of LLM decoding. By restricting the generation within a domain-specific context, the framework effectively narrows the generative search space, densely concentrating valid and highly accurate solutions within the top-$k$ candidates.

\begin{figure}
\centering
  \setlength{\abovecaptionskip}{1mm}
  \setlength{\belowcaptionskip}{-4mm}
\includegraphics[width=\linewidth]{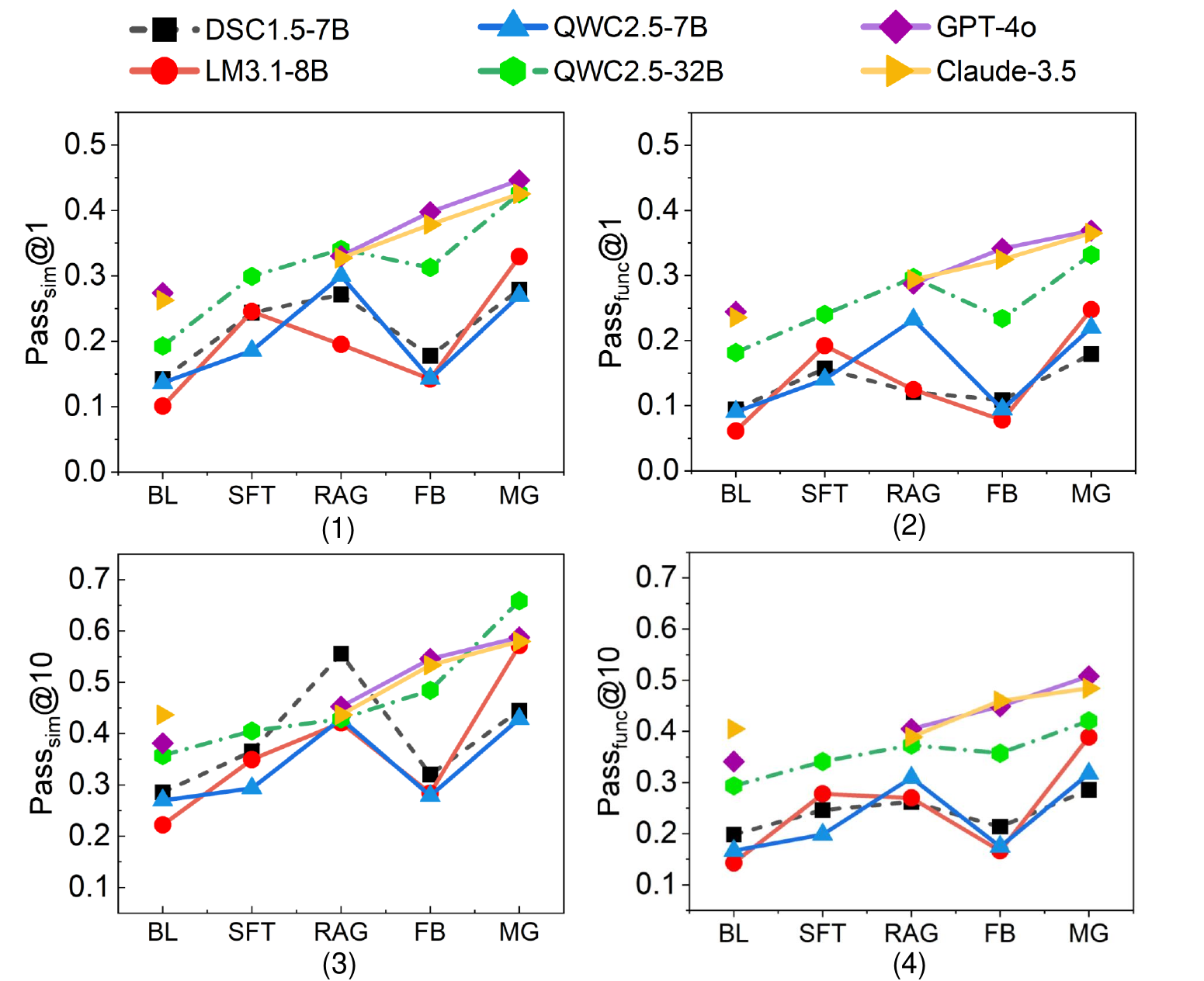}
\caption{Impact of Isolated Enhancement Strategies on Modelica Component Generation (BL = Baseline).}
\Description{}
\label{fig:5}
\end{figure}

\subsection{Evaluation of Enhancement Strategies}

To address \textbf{RQ2}, we evaluate the specific contributions of the proposed enhancement strategies. Figure \ref{fig:5} illustrates the isolated and combined effects of these strategies on Modelica component generation. To further validate the necessity of each module within the integrated framework, we conduct an ablation study with the corresponding results detailed in Figure \ref{fig:6}. We provide a comprehensive analysis of these results below.

\textbf{Impact of Domain-Specific Fine-Tuning.}
Figure \ref{fig:5} shows that domain-specific fine-tuning (\textbf{SFT}) serves as a crucial foundation for the open-source LLMs. When applied in isolation, SFT yields consistent gains in both the simulation and functional pass rates. For instance, compared to the baselines, the $pass_{sim}@1$ and $pass_{func}@1$ for LM3.1-8B increase from 10.08\% and 6.11\% to 24.52\% and 19.21\%, respectively. These consistent gains underscore that SFT successfully internalizes domain knowledge, leading to improvements in syntactic correctness, semantic accuracy, and physical consistency.

However, the ablation results in Figure \ref{fig:6} reveal a nuanced interaction between parametric memory and external retrieval. We observe a performance regression where, for QWC2.5-7B, using full ModiGen leads to a slight drop in $pass_{func}@1$ compared to the ablated state (\textbf{no\_SFT}). Notably, this phenomenon is even more pronounced for $pass_{sim}@k$, although $pass_{sim}@k$ does not represent the final outcome. For open-source LLMs, SFT builds structural patterns directly into the LLM's parameters, while RAG adds external constraints to the prompt. This observation demonstrates that forcing a fine-tuned LLM to process RAG prompts for every sample can disrupt its internalized knowledge, leading to a lower pass rate. This justifies the design of the \textit{conditional routing mechanism}, which ensures a steady improvement in overall performance.

\textbf{Effectiveness of Structural RAG.}
The structured RAG module (\textbf{RAG}) provides precise domain context for both proprietary and open-source LLMs. As illustrated in Figure \ref{fig:5}, the module consistently improves the evaluation metrics when integrated into the baselines. For instance, the $pass_{sim}@1$ for Claude-3.5 improves from 26.27\% to 32.70\%, while the $pass_{func}@1$ rises from 23.57\% to 29.37\%. The pivotal role of this module within the full framework is further validated by the ablation results in Figure \ref{fig:6}. Removing the retrieval component (\textbf{no\_RAG}) results in a stark performance decline across all LLMs. Notably, the $pass_{sim}@10$ for QWC2.5-7B drops from 42.86\% in the full framework to 30.16\%, and the $pass_{func}@1$ decreases from 21.98\% to 14.52\%. These results demonstrate that explicitly retrieving structural boundaries and component documentation effectively guides the generative process, mitigating semantic hallucinations and physical inconsistencies that often plague LLMs in physical modeling tasks.

Additionally, Table \ref{tab:3} highlights the varying marginal utility of the RAG module. We observe that the integration of RAG yields a much more pronounced improvement in open-source LLMs than in large proprietary LLMs. This indicates that while proprietary LLMs already internalize vast amounts of physical and syntactic knowledge, lighter open-source LLMs rely heavily on RAG to bridge this capability gap.

\begin{table}
\centering
  \setlength{\abovecaptionskip}{1mm}
  \setlength{\belowcaptionskip}{0mm}
\caption{\label{tab:3}Percentage improvement of $pass@1$ under different strategies.}
\scalebox{0.87}{
\setlength{\tabcolsep}{6pt}
\begin{tabular}{l|ccc|ccc}
\toprule
 & \multicolumn{3}{c|}{$\uparrow pass_{sim}@1 (\%)$} & \multicolumn{3}{c}{$\uparrow pass_{func}@1 (\%)$} \\
\cmidrule(lr){2-4}\cmidrule(lr){5-7}
LLM / Strategy & FB & RAG & MG & FB & RAG & MG \\
\midrule
DSC1.5-7B & 25.11 & 91.01 & 96.04 & 15.18 & 27.79 & 90.01 \\
LM3.1-8B & 40.94 & 93.69 & 226.75 & 27.30 & 103.93 & 305.27 \\
QWC2.5-7B & 4.66 & 119.78 & 97.69 & 4.36 & 156.95 & 142.92 \\
QWC2.5-32B & 62.14 & 76.54 & 120.99 & 28.96 & 63.50 & 82.73 \\
GPT-4o & 45.22 & 20.58 & 62.90 & 39.64 & 17.55 & 51.00 \\
Claude-3.5 & 44.11 & 24.47 & 61.93 & 37.72 & 24.59 & 54.89 \\
\bottomrule
\end{tabular}}
\end{table}

\begin{figure}
\centering
  \setlength{\abovecaptionskip}{1mm}
  \setlength{\belowcaptionskip}{-4mm}
\includegraphics[width=\linewidth]{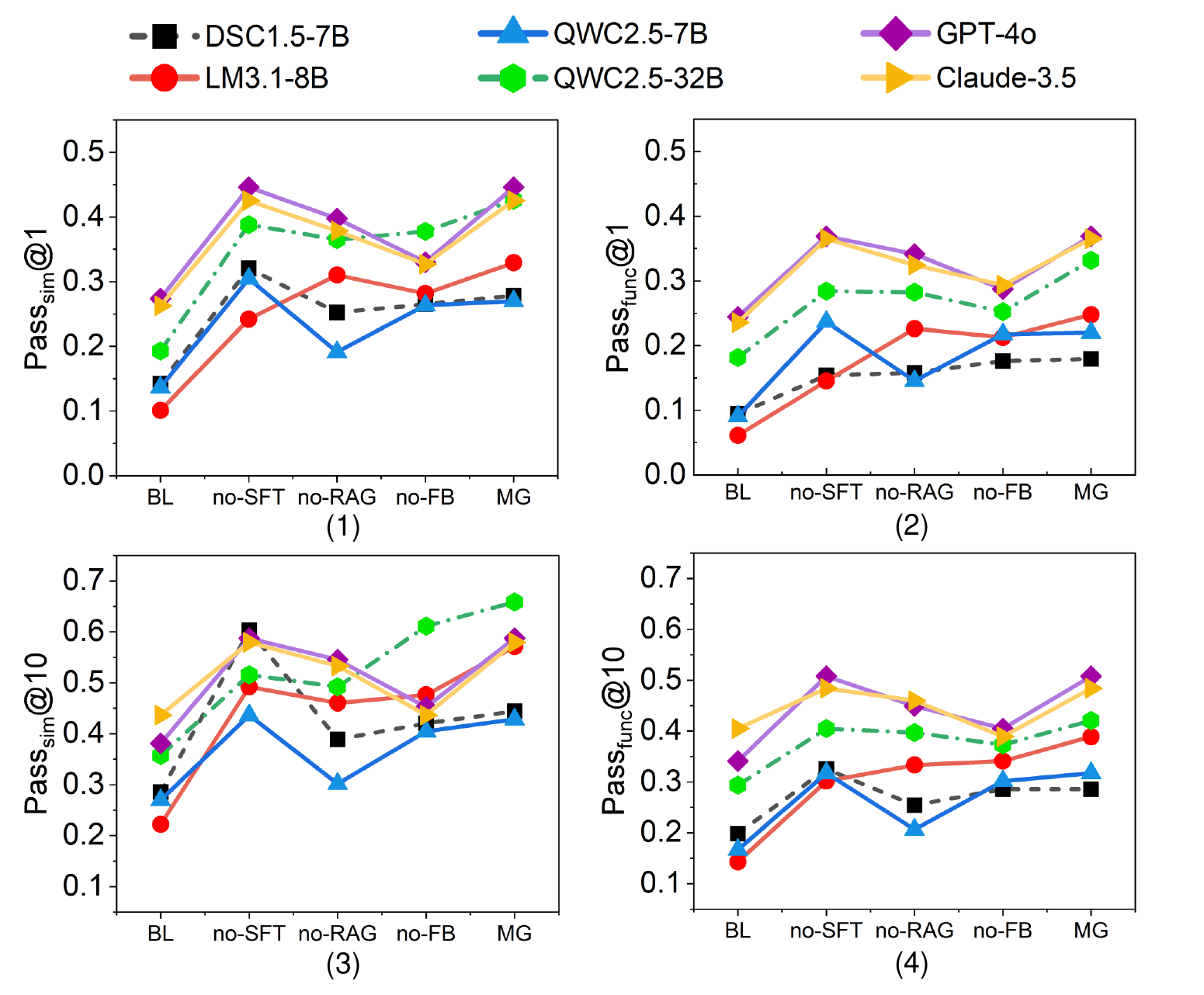}
\caption{Ablation Study of the ModiGen Framework.}
\Description{}
\label{fig:6}
\end{figure}

\textbf{Contribution of Feedback Refinement.}
Figure \ref{fig:5} and Figure \ref{fig:6} collectively highlight the role of feedback-driven refinement as a dynamic error-correction mechanism. In the component generation task, the inclusion of the feedback-driven refinement module (\textbf{FB}) allows LLMs to move beyond their initial outputs by rectifying detected flaws. For example, the baseline $pass_{func}@1$ for QWC2.5-32B increases from 18.15\% to 23.41\% with FB. Meanwhile, the ablation results in Figure \ref{fig:6} show that removing this module (\textbf{no\_FB}) leads to a consistent performance drop across the board, particularly evident on LM3.1-8B, GPT-4o, and Claude-3.5.

Notably, a comparison of $pass_{sim}@1$ and $pass_{func}@1$ improvements in Table \ref{tab:3} shows that feedback refinement yields slightly higher percentage gains in $pass_{sim}@1$ than in $pass_{func}@1$ for most LLMs. This disparity suggests that the feedback mechanism primarily enhances the syntactic and structural integrity of the generated Modelica component. Since the compiler-generated diagnostics focus heavily on resolving syntax violations and semantic errors in non-compliant components, the refinement process is particularly effective at guiding components toward successful simulation.

\begin{figure}
\centering
  \setlength{\abovecaptionskip}{1mm}
  \setlength{\belowcaptionskip}{-4mm}
\includegraphics[width=\linewidth]{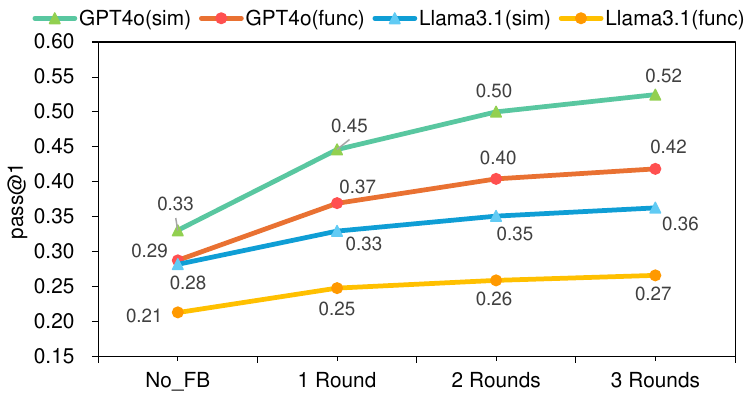}
\caption{Generation Performance of Representative LLMs across Multiple Feedback Rounds.}
\Description{}
\label{fig:7}
\end{figure}

In theory, multiple rounds of feedback optimization can continuously refine generation performance, as each iteration corrects validated errors based on execution feedback. To validate this, we selected representative LLMs (GPT-4o and LM3.1-8B) and evaluated their pass rates across 0 to 3 feedback rounds. As illustrated in Figure \ref{fig:7}, increasing the number of feedback rounds consistently improves generation performance, and the first round yields the most substantial performance leap across all configurations. However, this optimization exhibits diminishing marginal returns: as output quality increases in earlier iterations, the scope for further correction naturally shrinks. This is clearly evident in the subsequent rounds for $pass_{func}@1$ on LM3.1-8B, which yield progressively smaller increments, reaching 0.26 and 0.27 in rounds two and three, respectively. Given that the first round captures the vast majority of potential gains, we adopt a \textbf{single-round} feedback configuration in our framework to achieve an optimal trade-off between generation quality and computational overhead.

\textbf{LLM-Specific Module Dependency.} As evidenced by the results of $pass_{func}@k$ in Figure \ref{fig:5} and Figure \ref{fig:6}, the dependency on specific modules varies significantly even among open-source LLMs. For instance, the Qwen2.5-Coder series exhibits a more pronounced reliance on the RAG module, whereas LM3.1-8B benefits predominantly from SFT. This discrepancy suggests that code-specialized LLMs such as Qwen2.5-Coder already possess strong code reasoning and primarily require RAG to supply precise external topological constraints, while general-purpose LLMs such as LM3.1-8B must first rely on SFT to internalize the foundational syntax and domain-specific rules of Modelica. Ultimately, these complementary modules function synergistically, providing comprehensive guidance that drives the significant surges in evaluation metrics for these open-source LLMs.

\subsection{Overhead}
To address \textbf{RQ3}, we evaluate the computational overhead in terms of average latency and context tokens per component, as shown in Table \ref{tab:4}. Note that absolute latency depends on hardware specifications and serves as a relative baseline. While the baseline uses original pre-trained LLMs, the RAG, FB, and full ModiGen configurations utilize fine-tuned LLMs, which inherently maintain the same token scale as the baseline.
Results show RAG adds marginal latency but increases context tokens to accommodate structural documentation. Meanwhile, FB noticeably increases both latency and token usage due to compilation, dynamic simulation, diagnostics, and a secondary inference pass. Although ModiGen incurs the highest overall overhead, this investment remains highly cost-effective. Compared to hours of manual debugging or repeated zero-shot prompting, the computational cost of ModiGen is justified by its substantial gains in component generation pass rates.

\begin{table}
\centering
  \setlength{\abovecaptionskip}{1mm}
  \setlength{\belowcaptionskip}{0mm}
\caption{\label{tab:4}Average Time and Context Token Overhead per Component.}
\scalebox{0.84}{
\begin{tabular}{l|cccc|cccc}
\toprule
 & \multicolumn{4}{c|}{Time(s)} & \multicolumn{4}{c}{Total Tokens} \\
\cmidrule(lr){2-5}\cmidrule(lr){6-9}
Model/Strategy & BL & RAG & FB & MG & BL & RAG & FB & MG \\
\midrule
DSC1.5-7B & 7.05 & 9.17 & 13.15 & 15.87 & 989 & 1505 & 1732 & 2248\\
LM3.1-8B & 9.67 & 11.25 & 15.20 & 18.75 & 885 & 2151 & 1895 & 3161 \\
QWC2.5-7B & 7.68 & 9.64 & 14.92 & 16.88 & 779 & 1512 & 1434 & 2167\\
QWC2.5-32B & 12.71 & 16.77 & 26.38 & 31.32 & 873 & 1765 & 1702 & 2594\\
GPT-4o & 7.03 & 6.75 & 14.36 & 15.45 & 576 & 1573 & 1781 & 2778 \\
Claude-3.5& 7.16 & 7.34 & 16.10 & 16.23 & 679 & 1554 & 1994 & 2869 \\
\bottomrule
\end{tabular}}
\end{table}

\subsection{Case Study}
To illustrate the practical utility of ModiGen, we showcase two representative cases from the component generation task. 

\begin{figure}
\centering
  \setlength{\abovecaptionskip}{0mm}
  \setlength{\belowcaptionskip}{-4mm}
\includegraphics[width=\linewidth]{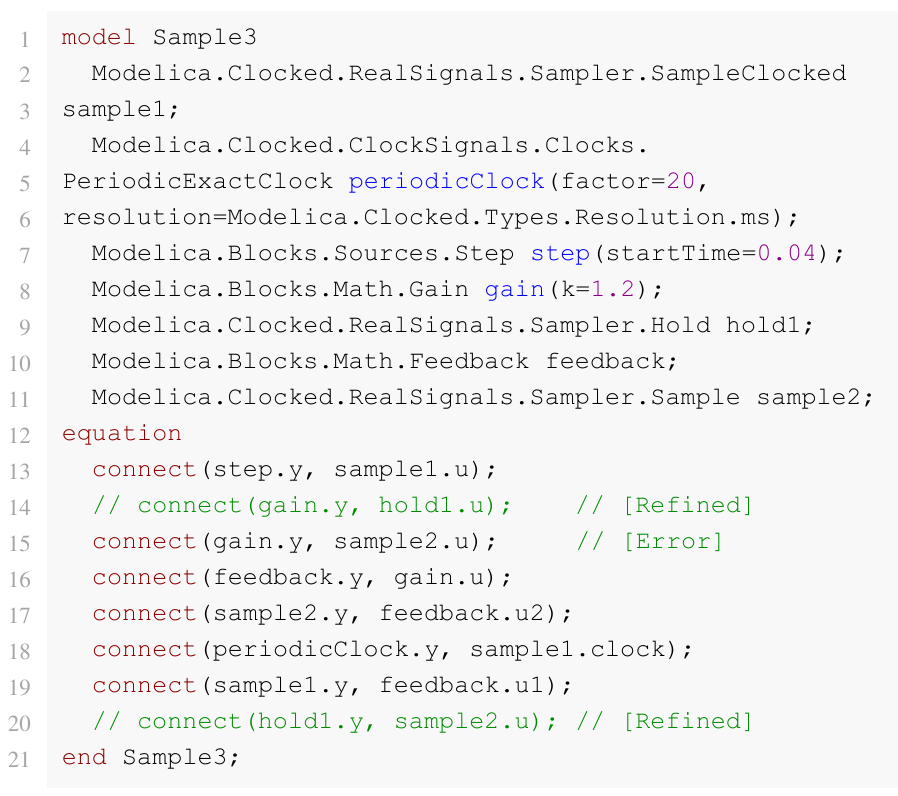}
\caption{Physical consistency for a clocked sample block.}
\Description{}
\label{lst:3}
\end{figure}

\textbf{Case I: Knowledge-Augmented Physical Consistency.} This case involves constructing a clocked sampling system in Figure~\ref{lst:3}, which features direct feed-through across both continuous-time and clocked partitions. The LLM exhibits topology hallucination: it fails to interface the continuous-time and clocked signals through the required sampling and hold blocks. Specifically, it attempts to directly short-circuit the continuous-time output of the \texttt{gain} block to the input of \texttt{sample2}: \texttt{connect(gain.y, sample2.u)} (line 15). By retrieving the metadata for the \texttt{Hold} block from the structural index, the LLM identifies the digital-to-analog boundary and recognizes that the continuous signal must be routed through the \texttt{Hold} block to serve as a valid input for \texttt{sample2}. Consequently, it correctly generates the \texttt{connect} statements (lines 14 and 20), ensuring the structural integrity of the sampled system. This highlights how knowledge augmentation anchors the generative process to the library structure, preventing topological inconsistencies that are common when LLMs deal with complex, multi-domain libraries.

\begin{figure}[h]
\centering
  \setlength{\abovecaptionskip}{0mm}
  \setlength{\belowcaptionskip}{-4mm}
\includegraphics[width=\linewidth]{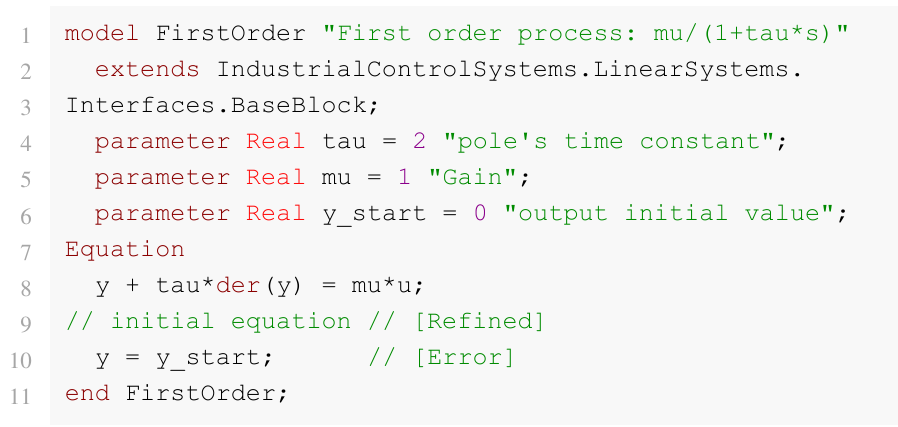}
\caption{Semantic refinement for the \texttt{FirstOrder} model.}
\Description{}
\label{lst:4}
\end{figure}

\textbf{Case II: Feedback-driven Semantic Refinement.} This case considers the generation of a first-order model. As illustrated in Figure~\ref{lst:4}, the initial output correctly formulated the differential equation \texttt{y + tau * der(y) = mu * u} but erroneously introduced a procedural initialization constraint \texttt{y = y\_start} (line 10) within the continuous \texttt{equation} section. This error occurs because the LLM confuses a one-time initial assignment with a continuous mathematical constraint. Influenced by procedural programming, it unintentionally forces the variable to remain constant, contradicting the differential equation. The verification module captured the following diagnostics: \textit{``[FirstOrder: 8:3-8:14]: Class FirstOrder has 2 equation(s) and 1 variable(s).''} Upon receiving this feedback, the refinement module prompted the LLM to relocate the initialization to the \texttt{initial equation} section (line 9). This case demonstrates how the feedback loop acts as an automated compliance layer, effectively mitigating semantic hallucinations and ensuring that generated components maintain algebraic completeness.



\section{Discussion} 
This section discusses our findings from three perspectives: prompt design, domain-specific adaptability, and system-level modeling, which clarify the practical conditions under which ModiGen can be effectively applied and extended beyond the evaluated benchmark.

\textbf{The Granularity of Prompts.}
The effectiveness of Modelica component generation heavily depends on prompt granularity. Prompts must provide sufficient detail to convey task requirements while avoiding excessive manual effort that could hinder processing efficiency \cite{white2023prompt}. In practice, prompts typically include task descriptions, modeling requirements, and parameters. For simpler tasks, coarser prompts may suffice. However, as task complexity increases—such as inferring topological relationships for components with more interfaces—LLMs often struggle to establish correct connections autonomously. Generally, well-structured and comprehensive prompts tend to enhance consistency and accuracy of generation. The prompt templates developed in this work (defined in Section \ref{sec:template}) offer guidance for designing prompts that effectively capture both task requirements and modeling constraints, without being overly prescriptive and thereby compromising the LLM's modeling autonomy.


\textbf{Domain-Specific Adaptability.} Although this study primarily evaluates ModiGen in general Modelica component generation, the framework is inherently appropriate for various kinds of domain-specific cyber--physical systems. Extending ModiGen to a new Modelica subdomain mainly requires two steps: preparing a domain-specific fine-tuning dataset that reflects the modeling conventions and semantic patterns of the target discipline, and organizing structured domain knowledge to support retrieval augmentation. For broader MDE domains beyond Modelica, the framework can be adapted by replacing its modular resources and interfaces, including target-language corpora for SFT and RAG, DSL-specific parsers or extractors, domain-specific verification tools, and error templates aligned with target compiler diagnostics. With these components in place, the supervised fine-tuning, RAG, and feedback-driven refinement modules can adapt to the domain’s modeling requirements. This modular design makes ModiGen a scalable framework for developing domain-specialized, LLM-powered modeling assistants.

\textbf{System-Level Perspective.} Current generation tasks primarily focus on producing components or small-scale subsystems, often lacking a holistic understanding of the overall system architecture. In complex engineering applications, a system-level perspective that captures the interactions and coordination among components is crucial. To generate coherent systems, LLMs can emulate the top-down design workflow of human domain experts: first conceptualizing the system architecture, then instantiating and verifying foundational components, and finally orchestrating their topological connections and multi-domain interactions. Future work could explore strategies such as role-based multi-agent collaboration \cite{hong2023metagpt}, hierarchical planning \cite{yao2023tree}, or graph-enhanced reasoning \cite{edge2024local} to strengthen the system-level reasoning ability of LLMs.

\section{Threats to validity}

\textbf{Internal Validity}. We address potential data contamination. A 3-gram Jaccard analysis between the benchmark components and $Dataset_{sft}$ shows an average similarity of 40.61\%, where the overlap mainly arises from standardized Modelica syntax, interface conventions, and recurring modeling patterns rather than identical functional designs. In addition, our dataset intentionally incorporates 21 author-synthesized components. Furthermore, the consistent underperformance of zero-shot baselines indicates that mere pre-training exposure is insufficient. The generalized gains achieved only through our framework suggest that success stems from our approach rather than reliance on data contamination.

\textbf{External Validity}. The diversity of the benchmark across critical physical domains ensures that the results are not case-specific. The results suggest that the framework can address common LLM-based modeling bottlenecks and has potential applicability across a broader range of engineering domains.

\textbf{Construct Validity}. Our evaluation relies on $pass_{sim}@k$ for simulation feasibility and $pass_{func}@k$ for functional correctness. While simulation reliably verifies syntax and topology, automated functional metrics might yield false positives due to numerical tolerances. To mitigate this, we supplemented the functional validation with rigorous manual inspection, ensuring the generated components adhere to the intended physical logic.

\section{Conclusion}
This study investigates the performance of large language models (LLMs) in generating Modelica component models. To advance this research, we developed a comprehensive benchmark that standardizes the evaluation of valid components. To improve generation quality, we propose an effective and adaptable framework that integrates fine-tuning, Modelica-specific structural retrieval-augmented generation, and feedback-driven refinement. Our evaluation across a diverse set of LLMs demonstrates that ModiGen effectively addresses key limitations, including syntactic errors, semantic hallucinations, and physical inconsistencies. Specifically, the proposed framework achieves significant performance gains, with improvements of up to 22.86\% in the simulation pass rate and 18.65\% in the functional validation pass rate. These results highlight the potential of LLMs in Modelica component generation and provide valuable insights into their capabilities in system modeling. Future research should focus on extending these component-level capabilities toward fully autonomous, system-level modeling, enabling LLMs to comprehend and construct complex cyber-physical architectures with minimal human intervention.

\begin{acks}
This work was partly supported by the National Natural Science Foundation of China under Grant No. 62302443 and by the Fundamental Research Funds for the Central Universities under Grant No. 2025ZFJH02.
\end{acks}

\bibliographystyle{ACM-Reference-Format} 
\bibliography{modigen}

\end{document}